\documentclass[letter,twocolumn,showpacs,aps,prl]{revtex4}

\expandafter\let\csname equation*\endcsname\relax
\expandafter\let\csname endequation*\endcsname\relax

\usepackage{amsmath}
\usepackage{amssymb}
\usepackage{color}
\usepackage{verbatim}
\usepackage{array}
\usepackage{graphicx}
\usepackage{epsfig}
\usepackage{bm}
\usepackage[ragged]{sidecap}
\usepackage{dsfont}
\usepackage{ulem}
\usepackage{mathrsfs}
\usepackage{setspace}

\usepackage{subfigure}

\everymath{\displaystyle}
\newcommand{\ds}{\displaystyle}
\newcommand{\ddsum}[1]{{\displaystyle \sum_{ #1 }}}

\newcommand{\supercomas}[1]{``#1''}

\newcommand{\ket}[1]{\mathinner{|{#1}\rangle}}

\newcommand{\ketbra}[2]{|#1\rangle \langle#2|}
\newcommand{\expval}[1]{\langle #1 \rangle}

\newcommand{\dd}{\mathrm{d}}

\newcommand{\Tr}[1]{\textrm{Tr}\left[#1\right]}

\begin{document}

\preprint{\hfill\parbox[b]{0.3\hsize}{ }}

\title{
Full-quantum light diode
}

\author{F. Fratini\footnote{
\begin{tabular}{ll}
E-mail addresses:& filippo.fratini@tuwien.ac.at\\
&fratini.filippo@gmail.com
\end{tabular}}
and R. Ghobadi
}
\affiliation
{\it
Atominstitut, TU Wien, Vienna, Austria
}

\date{\today}

\begin{abstract}
Unidirectional light transport in one-dimensional nanomaterials at the quantum level is a crucial goal to achieve for upcoming computational devices. We here employ a full-quantum mechanical approach based on master equation to describe unidirectional light transport through a pair of two-level systems coupled to a one-dimensional waveguide. By comparing with published semi-classical results, we find that the nonlinearity of the system is reduced, thereby reducing also the unidirectional light transport efficiency. Albeit not fully efficient, we find that the considered quantum system can work as a light diode with an efficiency of $\approx 60$\%. Our results may be used in quantum computation with classical and quantized light.
\end{abstract}


\maketitle

{\it Introduction:}
One-dimensional (1D) systems are emerging as promising candidates for a new generation of computational devices. The idea leading this research field is to build a quantum interface between light and matter, so as to replace (or hybridize) electronic devices by optoelectronic devices in the next future \cite{Warburton2008, Kimble2008}. Light based devices outclass electronic devices in terms of low energy dissipation, large data transfer, high quantum information control, and robustness against environmental decoherence. 
During the last decade experiments have succeeded in coupling single quantum emitters to 1D systems, in a variety of well-established technologies, such as superconducting circuits, semiconductor quantum dots, and nitrogen vacancy centers in diamonds \cite{Huck2014, Astafiev2010, VanLoo2013, Abdumalikov2010}. Likewise, theoretical studies have been directed toward conceiving basic optoelectronic devices that are able to work at the single- or few- photon level, such as quantum optical diodes \cite{Roy2010, Shen2011, Mascarenhas2014, Shen2014, Fratini2014}. 

\begin{figure}[t!]
\begin{center}
\includegraphics[scale=0.35, trim=0cm 0cm 0cm 0cm, clip=true]{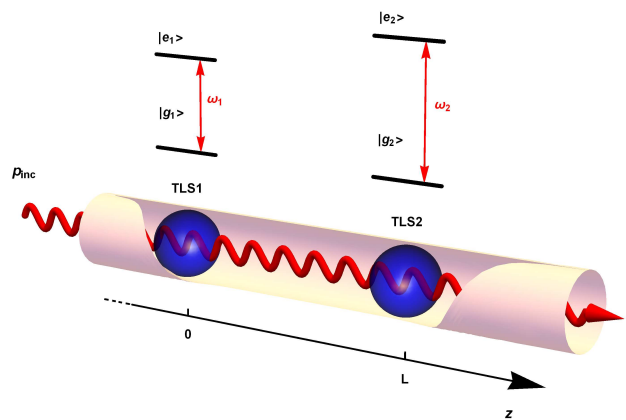}
\caption{(color online). The system we consider: A pair of two-level systems that is embedded in a one-dimensional waveguide, and that is irradiated by coherent light.
}
\label{fig:Setting}
\end{center}
\end{figure}

Optical diodes are devices that permit unidirectional light transport in 1D systems. Very recently there has been a research boost in looking for unidirectionality in 1D systems. Several experiments have reported unidirectional light propagation in metamaterials \cite{Feng2013, Estep2014, Neugebauer2014, Fortuno2013, Feber2015} and polarization dependent light transport (chirality) in nanofibers \cite{Mitsch2014, Petersen2014}. These works are motivated by the fact that, as electronic diodes have been responsible for the electronic revolution of the last century, photonic diodes are expected to play a crucial role in the development of optoelectronics. With this goal in mind, we here present a full-quantum (FQ) analysis of the unidirectional capabilities of pair of two-level systems (PTLSs) coupled to a 1D waveguide, as depicted in Fig. \ref{fig:Setting}. A PTLSs is the simplest 1D configuration able to manifest tunable nonlinearity at the quantum level. Such nonlinearities are fundamental for exploiting nonreciprocal light transport \cite{Khanikaev2015}. Thus, a PRLSs represents a valid candidate for the essential building block to control light transport at the quantum level for next generation light based devices.

Our results are compared with a recent work of one of us, where a semi-classical (SC) model has been employed to test the nonreciprocal capabilities of the PTLSs \cite{Fratini2014}.
We show that the PTLSs does work as a light diode for certain values of detunings and inter-atomic distance, confirming the SC results. However, we find that the FQ description predicts lower nonlinear behaviour of the system compared with the SC description. Light rectification, which emerges out of the TLSs nonlinear response to light, turns out to be also reduced. 
Yet, the device is efficient enough to be used as quantum diode: In favorable settings, light transmittance through the PTLSs is found to be $\approx 0$\% from one side, while $\approx 60$\% from the other side. 
This work represents a conclusive analysis of the unidirectional light transport (rectification) capabilities of the PTLSs.

\medskip

{\it System:} 
The PTLSs we consider is shown in Fig. \ref{fig:Setting}. It consists of a pair of two level systems (TLSs) confined in a 1D waveguide. The first (TLS1) and second (TLS2) quantum emitters sit at positions $z_1=0$ and $z_2=L$, respectively. 
The TLS1 (TLS2) has transition frequency $\omega_{1(2)}$ and decay rate $\gamma_{1(2)}$. Light with angular frequency $\omega_0$ and wave-length $\lambda$ is injected into the waveguide. The detuning of the light with respect to the TLS1 (TLS2) transition frequency is $\delta\omega_{1(2)}=\omega_0-\omega_{1(2)}$. We suppose that the incident light be laser light (i.e., it is a coherent quantum state) whose power is measured by the number of photons per atomic lifetime ($p_{inc}$). In the following, for simplicity we shall consider equal atomic decay rates $\gamma_1=\gamma_2=\gamma=1$.

\medskip

{\it Theory:} 
The light propagation through the system can be described in quantum mechanics via a master equation \cite{Carmichael}. By using such an approach, the degrees of freedom of the quantized light can be traced out of the equation.  We are thus left with an equation describing only the atomic density operator ($\rho$). For a coherent driving light field and within the Markov approximation, the master equation reads \cite{Carmichael}
\begin{equation}
\begin{array}{lcl}
\dot\rho&=&-i\big[ H,\rho \big]+\ddsum{i,j=1,2}\frac{\gamma_{ij}}{2}
\Big(
2\sigma_{i-}\rho\sigma_{j+}\\
&&\quad -~ \sigma_{i+}\sigma_{j-}\rho~ -~ \rho\sigma_{i+}\sigma_{j-}
\Big)
\end{array}
\label{eq:rhodot}
\end{equation}
where
\begin{equation}
\begin{array}{lcl}
H&=&\ddsum{j=1,2}\left[
\omega_{j}\sigma_{j+}\sigma_{j-}-\frac{i}{2}\Big(
\Omega_j(t)\sigma_{j+}-\Omega_j(t)^*\sigma_{j-}\Big)\right.\\
&&\quad\left. +
\ddsum{k\neq j}\frac{D_{jk}}{2}\big(
\sigma_{j-}\sigma_{k+}+\sigma_{j+}\sigma_{k-}
\big)
\right]~,\\[0.5cm]
\Omega_j(t)&=&\Omega e^{-i\omega_0(t-z_j/c)}~,\qquad \Omega=\gamma\sqrt{2p_{inc}\beta}\\
\gamma_{ij}&=&\gamma\cos\big[ \omega_0/c(z_i-z_j) \big] + \gamma_{bg}\delta_{ij}~,\\
D_{ij}&=&\frac{\gamma}{2}\sin\big[ \omega_0/c|z_i-z_j| \big]~,
\end{array}
\end{equation}
$\beta$ is the efficiency parameter \cite{Fratini2014}, $c=\omega_0\lambda/(2\pi)$ is the speed of light in the waveguide, $\gamma_{bg}$ is the background decay rate, $p_{inc}$ is the incident intensity (photons per atomic lifetime), and we considered $\hbar=1$. Standard notation is used for atomic operators: $\sigma_{j+}=\ketbra{e_j}{g_j}$, $\sigma_{j-}=\ketbra{g_j}{e_j}$, $\sigma_{jz}=\ketbra{e_j}{e_j}-\ketbra{g_j}{g_j}$.  Electric dipole and rotating wave approximations have been used. As we are not interested in inefficiencies in the present work, we will assume perfect efficiency throughout the article ($\beta=1$). We will furthermore assume a small background decay rate ($\gamma_{bg}\approx 10^{-4}$).

The slowly varying atomic operators for the system ($S$-operators) can be defined as $S_{j\mp}=e^{\pm i\omega_0(t-z_j/c)}\sigma_{i\mp}$, and $S_{jz}=\sigma_{jz}$, where $j=1,2$. There are here nine independent $S$-operators: $S_{1-}$, $S_{1z}$, $S_{2-}$, $S_{2z}$, $S_{1-}S_{2z}$, $S_{1-}S_{2-}$, $S_{1z}S_{2-}$, $S_{1z}S_{2z}$, $S_{1-}S_{2+}$. 
The Bloch-Langevin equations for the expectation values over such operators are obtained via the relation $\expval{\dot S}=\Tr{\dot\rho S}$, where $S$ is any of the atomic $S$-operators. If we restrict the summation to $j=1$, the system of equations reduces to the well-known equations for a single TLS in the 1D waveguide \cite{Valente2012}. The full system of nine equations has been here automatically resolved by using the $Quantum$ Mathematica package \cite{Gomez2013}. As we are interested in the time independent response of the system to light, we looked for solutions of the expectation values of the $S$-operators in the steady-state regime. 
At the first order in $\Omega$, simple solutions for $S_{1-}$ and $S_{2-}$ can be obtained:
\begin{equation}
\begin{array}{lcl}
\expval{S_{1-}}^{(1)}&=&-\frac{\Omega}{2}\frac{\alpha_2-\beta_1}{\alpha_1\alpha_2-\beta_1\beta_2}~,\\[0.4cm]
\expval{S_{2-}}^{(1)}&=&\left.\expval{S_{1-}}^{(1)}\right|_{1\leftrightarrow 2}~,
\end{array}
\label{eq:SSolApprox}
\end{equation}
where ${}^{(1)}$ indicates the first order approximation, while $\alpha_{1(2)}=\gamma/2-i\delta\omega_{1(2)}$ and $\beta_{1(2)}=e^{-i\frac{\omega_0}{c}(z_{1(2)}-z_{2(1)}} (\gamma_{12}/ 2+iD_{12})$.

\medskip

In order to analyze light transmission and reflection, we need to analyze the electric field operator. Such an operator can be written at any point $z$ and time $t$ in terms of atomic $S$-operators as
\begin{equation}
\begin{array}{lcl}
E_F(z,t)&=&E_{F,in}(z,t)+\sqrt{\frac{\gamma}{2}}e^{-i\omega_0(t-z/c)}\ddsum{j=1,2}S_{j-}\theta(z-z_j)~,\\
E_B(z,t)&=&\sqrt{\frac{\gamma}{2}}e^{-i\omega_0(t+z/c)}\ddsum{j=1,2}S_{j-}\theta(z_j-z)
\end{array}
\label{eq:Efield}
\end{equation}
where F (B) stand for forward (backward) propagating light, $E_{F,in}$ is the incident electric field operator, while $\theta(z)$ is the step-function. The operator $E_{F,in}$ satisfies the relation $E_{F,in}\ket{\alpha}=\sqrt{\gamma p_{inc}}\,e^{-i\omega_0(t-z/c)}\ket{\alpha}$, where $\ket{\alpha}$ is the incident coherent state. 

Quantities of interest are obtained through the expectation value of different combinations of the electric field operator, where the expectation value is taken over the input state. For instance, the fraction of light power that is transmitted through the system, i.e. the transmittance, is obtained as $T=\expval{E^\dagger_F (z,t) E_F (z,t)}/\expval{E_{F,in}^\dagger(z,t)E_{F,in}(z,t)}$, where $z>L$.
Similarly, the fraction of transmitted field (transmission coefficient) is obtained as 
$t_k=\expval{E_F (z,t)}/\expval{E_{F,in}(z,t)}$, for $z>L$, which leads to 
$\textstyle t_k=\big(
1+\gamma\sum_j\expval{S_{j-}}/\Omega
\big)
$.

\medskip

{\it Linearity limitations of the system:}
The transmission coefficient at the first order in $\Omega$ is given by
\begin{equation}
\begin{array}{l}
t_k^{(1)}\equiv \left. t_k\right|_{S_{j-}\to S_{j-}^{(1)}}\\
\qquad=\frac{\delta\omega_1\delta\omega_2}{
\left(\delta\omega_1+i\frac{\gamma}{2}\right)
\left(\delta\omega_2+i\frac{\gamma}{2}\right)
-\frac{\gamma^2}{4}e^{2i\omega_0 L/c}
}~,
\end{array}
\label{eq:t_Linear}
\end{equation}
and coincides with the transmission coefficient found for the scattering of a single photon by the PTLSs \cite{Zheng2013}. This derivation shows that, in 1D settings, an incident coherent field in the limit $p_{inc}\to 0$ is equivalent to an incident single photon, as far as field transmission (both amplitude and phase) is concerned. From Eq. \eqref{eq:t_Linear}, one can also see that the transmitted field ratio $t_k^{(1)}$ is $\delta\omega_1\leftrightarrow\delta\omega_2$ symmetric, thereby yielding zero light rectification. This in turn shows that, in the limit of low incident intensity, the system becomes linear, thus no light rectification can be obtained. 

\begin{figure}[b]
\centering
\includegraphics[scale=0.5]{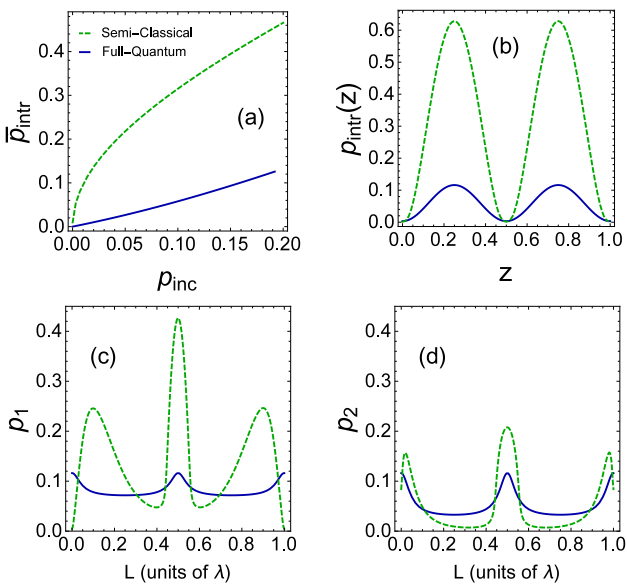}
\caption{(color online). 
(a) Average intracavity intensity; (b) Intracavity intensity;
(c) Intensity impinging onto TLS1; (d) Intensity impinging onto TLS2. 
$p_{inc}$ is the incident power, while $L$ is the TLS distance in units of photon wavelength. 
$\delta\omega_1=\delta\omega_2=0$ in all panels. 
$L=1$ in (a) and (b). $p_{inc}=0.1$ in (b), (c) and (d).
}
\label{fig:Fig2}
\end{figure}

The linear and nonlinear response of the PTLSs can be best studied by analyzing the light intensity in between the atoms (intracavity light intensity). To this aim, let us denote by $p_{intr}(z)=\expval{|E_F(z,t)+E_B(z,t)|^2}$ the intracavity light intensity at the position $z$, where $0<z<L$, and by $ \overline{p}_{intr}$ the average intracavity light intensity: $\overline{p}_{intr}=\int_0^L p_{intr}(z) \dd z/L$.
In Figs. \ref{fig:Fig2}(a) and (b), these two quantities are plotted as a function of $p_{inc}$ and $z$, respectively. The comparison with the SC results  \cite{Fratini2014} is shown. From the panel (a) one can see that, for $p_{inc}\lesssim 0.2$, the average intracavity intensity obtained in the FQ approach is considerably lower than the incident intensity, and is almost linear in $p_{inc}$. This marks a stark difference with the SC results, where at low incident intensity the average intracavity intensity is found to be nonlinear, much bigger than the incident intensity, and approximately equal to $\sqrt{p_{inc}}$. One may therefore conclude that the quantumness reduces the nonlinearity of the system, thereby setting limitations on the nonlinear effects that can be exploited at the quantum level \cite{Khanikaev2015}. We shall in fact see that this leads to a reduction of light rectification. 

From Fig. \ref{fig:Fig2} (b), one sees that nodes of the field are found at positions $z=0$, $0.5$ and $1$, both in the FQ and SC regime. The intracavity field close to the atomic positions is thus approximately vanishing, as one would expect for a cavity with almost perfect mirrors. On the other hand, the intracavity intensity peaks are found to be reduced in the FQ approach with respect to the SC approach.
Finally, although not shown, $\overline{p}_{intr}$ can be well approximated by $\overline{p}_{intr}\approx p_{inc}$ for large $p_{inc}$, as expected.

In Figs. \ref{fig:Fig2}(c) and (d), we plot the light intensity at the sites of the TLSs. Their definition, in the FQ approach, is $p_1(z)=\expval{|E_F(0^-,t)+E_B(0^+,t)|^2}$ and $p_2(z)=\expval{|E_F(L^-,t)|^2}$, where ${x}^\pm$ means evaluated in the limit $z\to x^\pm$. The comparison with the SC results is shown. Although the resulting FQ and SC curves have similar shapes, they are quantitatively different. Such a difference may be considered a measurable non-classical feature of the system. 

\medskip

\begin{figure}[t]
\includegraphics[scale=0.4]{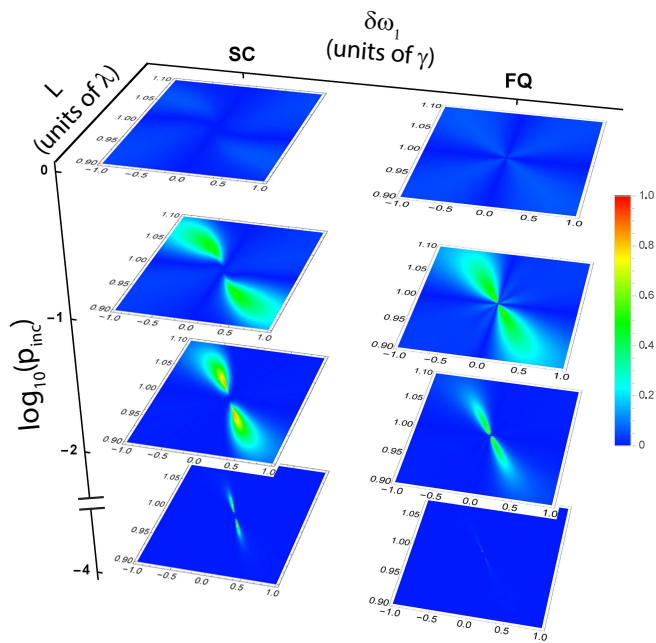}
\caption{(color online). The rectification efficiency parameter, $\mathcal{L}$, as obtained from the SC \cite{Fratini2014} and the FQ approaches. Different layers are related to different incident intensities $p_{inc}$, while $\delta\omega_2\approx 0$. $L$ is in units of photon wavelength, while $\delta\omega_1$ is in units of atomic decay rate.
}
\label{fig:LRC}
\end{figure}

{\it Light rectification:}
In order to explore the unidirectional light transport capabilities of the system, we define the \supercomas{rectifying factor} as \cite{Mascarenhas2014, Fratini2014}
\begin{equation}
\mathcal{R}=\ds\frac{\big| T_{12}-T_{21} \big|}{T_{12}+T_{21}}~,
\label{eq:RectFactor}
\end{equation}
where $T_{12}$ is the transmittance $T$ for the case light is shined from first-to-second atom in the waveguide (as in Fig. \ref{fig:Setting}), while $T_{21}$ is the transmittance in the optical inverse situation where light is shined from second-to-first atom. We shall take $\mathcal{L}=T_{12}\mathcal{R}$ as the figure of merit to quantify the \supercomas{rectification efficiency} of the PTLSs. The quantity $\mathcal{L}$ is shown in Fig. \ref{fig:LRC}, as obtained from the SC (left pile) and the FQ (right pile) approach, for the case $\delta\omega_2\approx 0$. 
$\mathcal{L}$ disappears at very high and very low incident light intensity. This defines an optimal incident intensity $p_{nc}$ by which to exploit unidirectional light transport through the PTLSs. Such an optimal value is found within the interval $10^{-2}\lesssim p_{inc}\lesssim 10^{-1}$. In the best areas, the FQ approach gives $\mathcal{R}\approx1$ and $\mathcal{L}\approx 0.6$, as largest values.
We conclude that, although not perfectly efficient, the PTLSs works as a relatively good light diode in favorable settings, transmitting $\approx0$\% light in the path 2-to-1, and $\approx 60\%$ light in the path 1-to-2. 

\begin{figure}[t]
\begin{center}
\includegraphics[scale=0.6]{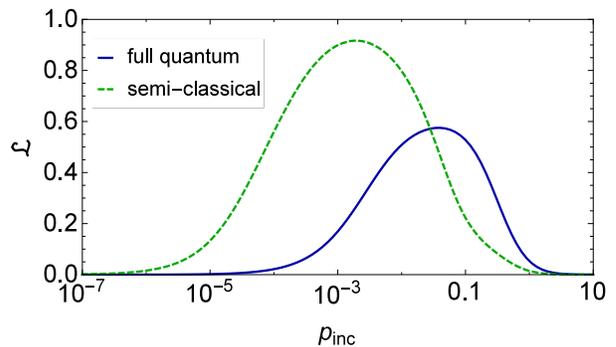}
\caption{(color online). The rectification efficiency parameter, $\mathcal{L}$, as obtained from the SC \cite{Fratini2014} and the FQ approach. Parameters $(\delta\omega_1, \delta\omega_2, L)$ are set as $(0.12, 0, 0.982)$. The plot shows that the rectification efficiency in the FQ description is reduced with respect to the SC description.
}
\label{fig:SemiFull}
\end{center}
\end{figure}

Qualitatively similar results for $\mathcal{L}$ are obtained semi-classically, as evident from Fig. \ref{fig:LRC}. However, there is a quantitative difference that is best evident in Fig. \ref{fig:SemiFull}, where $\mathcal{L}$ is plotted for one set of parameters $(\delta\omega_1, \delta\omega_2, L)$ yielding high light rectification. 
We see that the quantumness of the PTLSs reduces the rectification efficiency peak, and shifts it at slightly higher incident powers. Since light rectification requires nonlinearity, this reduction could be traced back to the aforementioned reduction of the intracavity nonlinearity. The difference between SC and FQ as in Figs. \ref{fig:LRC} and \ref{fig:SemiFull} can be thus also considered a measure of the non-classicality of the system. 
Such a difference is not wholly unexpected in virtue of the fact that
$\mathcal{L}$ takes large values in regions close to the single photon regime, where
the quantum nature of light is expected to be important.

\medskip

{\it Physical implementation and applications: }
The PTLSs can be implemented in several well-established technologies. Microwave superconducting circuits, for instance, would be a good testing bed for ascertaining the rectification properties presented here. This is because the coupling of such circuits to the 1D geometry is nearly 100\% \cite{VanLoo2013}. Another interesting platform where to implement the PTLSs are semiconductor quantum dots coupled to 1D photonic wires \cite{Gr1}. Such artificial atoms behave like ideal two-level quantum systems, and have been recently showed to function even at room temperature \cite{Bounouar2012, Holmes2014}. Nitrogen vacancy (N-V) centers can also be used. They are among the most stable and robust quantum emitters, due to the diamond shell that protects them from the environment. Their optoelectronic usefulness has been recently demonstrated, by employing them to build an optical switch \cite{Geiselmann2013}. Finally, nanoscale particles trapped in regions close to the waveguide can behave as PTLSs and have been showed to interact with the evanescent field of the injected light. Along the lines of this work, polarization dependent transport using such systems has been recently experimentally demonstrated \cite{Mitsch2014, Petersen2014}.

$n$-bit gate unitary operations are generally conceived in 1D settings. Within the standard Feynman notation for designing quantum gates, the time as well as the processing advances from left to right. This inevitable requires unidirectional transport of the q-bits \cite{Barenco1995}. To this aim, a photonic diode that is able to work at the quantum regime and that is integrable in 1D nano-materials, as presented in this Letter, can be very useful as a tool to suppress unwanted reflected noise. Such a device could be employed in quantum computation performed with quantized \cite{Knill2000} and with classical light \cite{Garcia2015}.

\medskip

{\it Acknowledgements: }
F.F. and R.G. acknowledge support from the Austrian Science Fund (FWF) through the START Grant No. Y591-N16.
F.F. acknowledges Peter Rabl, Laleh Safari, Dario Gerace and Eduardo Mascarenhas for fruitful discussions.


\end{document}